
\font\teneufm=eufm10 scaled \magstep1
\newfam\eufmfam
\textfont\eufmfam=\teneufm
\def\frak#1{\fam\eufmfam#1}
\magnification = \magstep1
\global\count0=0
\output={\plainoutput}
\def\plainoutput{\shipout\vbox{\makeheadline\pagebody\makefootline}%
\advancepageno
}
\footline={\ifnum\pageno=0\hss\else\hss\tenrm\folio\hss\fi}
\hsize=12.6cm
\overfullrule=0pt

\noindent
\null\hskip10cm MRR-009-92 \hfill\break
\null\hskip10cm SMS-042-92 \hfill\break
\null\hskip10cm ANU preprint \hfill\break
\null\hskip10cm May, 1992 \hfill\break
\vskip1cm
\centerline{\bf Spectra in Conformal Field Theories from the Rogers
Dilogarithm}\par
\vskip1cm
\noindent
\centerline{by}\par
\vskip1cm
\centerline{{\bf A. KUNIBA}\footnote\dag{Postal address:
 Department of Mathematics, Kyusyu University, Fukuoka 812}
 \footnote\null{E-mail: kuniba@math.sci.kyushu-u.ac.jp}
{\bf and T. NAKANISHI}\footnote\ddag{E-mail:
nakanisi@rabbit.math.nagoya-u.ac.jp}}
\vskip1.5cm
\noindent
\hskip1.2cm \dag\hskip0.2cm Mathematics and Theoretical Physics, \par\noindent
\hskip1.6cm Institute of Advanced Studies, \par \noindent
\hskip1.6cm Australian National University,\par\noindent
\hskip1.6cm  GPO Box 4 Canberra ACT 2601 Australia\par\noindent
\hskip1.6cm (On leave from Department of Mathematics, Kyushu University,\par
\noindent\hskip1.8cm Fukuoka 812 Japan)\par
\vskip0.4cm
\noindent
\hskip1.2cm \ddag \hskip0.2cm Department of Mathematics,
Nagoya University,\par\noindent
\hskip1.6cm Chikusa-ku, Nagoya 464 Japan\par\noindent
\vskip0.4cm\noindent
\vskip1.9cm\noindent
\centerline{\bf ABSTRACT}\par\vskip0.2cm
We propose a system of functional relations having a
universal form connected to the
$U_q(X^{(1)}_r)$ Bethe ansatz equation.
Based on the analysis of it, we conjecture
a new sum formula for the Rogers dilogarithm function in
terms of the scaling dimensions of the $X^{(1)}_r$ parafermion
conformal field theory.
\vfill\eject

\baselineskip0.6cm
\def\xr{X_r}
\def\axr{X^{(1)}_r}

\def\uqaxr{U_q(X^{(1)}_r)}
\def\Log{{\rm Log}}
\beginsection 1. Introduction

It has been noticed for some time that
the Rogers dilogarithm function curiously emerges in the calculation of
physical quantities in various integrable systems.
Besides the outstanding result in three dimensions [1],
one finds extensive examples in two dimensions such as in
finite size corrections to the ground state energy
[2,3],
low-temperature asymptotics of the specific heat capacity
[4-8] and
ultraviolet Casimir energies [9-13]
in perturbed conformal field theories (CFTs) [14].
All these quantities are essentially identifiable with the
central charges of the relevant CFTs [4,15]
and have often been evaluated
in terms of the Rogers dilogarithm function [16].
It is therefore tempting to seek a universal background for the
common appearance of
the dilogarithm in the various calculational techniques
exploited in these works, e.g.,
functional relations (FRs) among row-to-row transfer
matrices, string hypotheses, thermodynamic Bethe ansatz (TBA) [17] and so
forth.
\par
This Letter is our first step toward understanding
the interrelation among these approaches.
We shall propose a
FR having a universal form related to the
$\uqaxr$ Bethe ansatz equation [18].
It may be viewed as a ``spectral parameter" dependent version of the recursion
relation among the ``Yangian characters" [19] whose combinations appear in
the arguments of the dilogarithm [7,8,20].
When $q = $ exp$({2\pi i \over \ell + g})$, we especially consider
a restricted version of the FR which closes among finitely many unknown
functions.
Here $\ell$ is a positive integer and
$g$ is the dual Coxeter number.
The restricted FR originates in the
equilibrium condition on the ratio of the string and hole densities in the
thermodynamic analysis of the $\uqaxr$ Bethe equation in [7,8].
On the other hand, the special case $\axr = A^{(1)}_1$ of the FR possesses
essentially the same form as that for certain combinations of
the actual transfer matrices [3,6] in
the restricted solid-on-solid (RSOS) models [21].
Although the full content of the FR is yet under investigation,
our heuristic analysis extending the $A^{(1)}_1$ case [3] leads to a new
conjecture that connects an analytically continued
Rogers dilogarithm function and the
scaling dimensions in parafermion CFTs [22].
It is basically labeled by a triad $(\axr, \ell, \Lambda)$ where
$\Lambda$ is the classical part of a level $\ell$ dominant integral weight of
$\axr$.
The earlier conjecture [20] (see also [8]) involving the central charges
concerns the special case $(\axr, \ell, 0)$ in our new one.
We expect further
extensions should be possible related to various coset CFTs by using
our FR.
The common structure of the FRs between the RSOS transfer matrices and
the TBA has also been observed in [3,23] for $\axr = A^{(1)}_1$.
\beginsection 2. $\uqaxr$ functional relation

Let $\axr$ denote one of the rank $r$ non-twisted affine Lie algebras [24]
$A^{(1)}_r (r \ge 1),
B^{(1)}_r (r \ge 2), C^{(1)}_r (r \ge 1), D^{(1)}_r (r \ge 3),
E^{(1)}_6, E^{(1)}_7,
E^{(1)}_8, F^{(1)}_4$ and $G^{(1)}_2$ and
$\uqaxr$ be the $q-$deformation of
its universal enveloping algebra [25].
Let $\alpha_a, \, \Lambda_a$ $(1 \le a \le r)$ be the simple roots and the
fundamental weights of the classical part $\xr$, respectively.
We set
$\rho = \sum_{a=1}^r \Lambda_a,\,
Q = \sum_{a=1}^r{\bf Z}\alpha_a,\,
P = \sum_{a=1}^r {\bf Z}\Lambda_a,\,
P_\ell = \{\Lambda \in P \vert
0 \le (\Lambda \vert \hbox{maximal root}) \le \ell \},\,
{\frak H}^\ast = \sum_{a=1}^r {\bf C}\Lambda_a$ and
${\frak H}^\ast_{\bf R} = \sum_{a=1}^r {\bf R}\Lambda_a$.
Introduce the bilinear form $(\, \vert \, )$ on ${\frak H}^\ast$ in
a standard way and normalize the roots as $\vert$ long root $\vert^2 = 2$.
Fix an integer $\ell \ge 1$ and put $\ell_a = t_a\ell, \,
G = \{ (a,m) \vert 1 \le a \le r,\, 1 \le m \le \ell_a - 1 \}$ throughout.
The following notation will be used in the sequel $(1 \le a, b \le r)$
$$\eqalignno{
&t_a = {2 \over (\alpha_a \vert \alpha_a) }, \quad
t_{a b} = \hbox{max}(t_a, t_b),&(1\rm a)\cr
&A^{m\, k}_{a\, b} = t_{a b}\bigl(
{\rm min}({m \over t_a}, {k \over t_b}) - {m k \over t_a t_b \ell} \bigr)
\quad\,{\rm for }\,\, (a,m), (b,k) \in G,
&(1\rm b)\cr
&B_{a b} = {t_b \over t_{a b}} C_{a b},\quad
C_{a b} = {2(\alpha_a \vert \alpha_b) \over (\alpha_a \vert \alpha_a) },
\quad I_{a b} = 2\delta_{a b} - B_{a b}.
&(1\rm c)\cr}
$$
Here $C_{a b}$ and
$B_{a b} = B_{b a}$ are the Cartan and the symmetrized Cartan matrix,
respectively.
The nodes in the Dynkin diagrams are numerated according to [24].
Let
$Y^{(a)}_m(u),\, 1 \le a \le r,\, m \ge 1$ be functions of a
complex parameter $u$ obeying
$$\eqalignno{
&Y^{(a)}_m(u+{i \over t_a})Y^{(a)}_m(u-{i \over t_a}) =
{\prod_{b=1}^r\prod_{k=1}^3 F_k^{I_{a b}\delta_{t_a k, t_{a b}}}
\over
\bigl(1+Y_{m-1}^{(a)}(u)^{-1}\bigr)\bigl(1+Y_{m+1}^{(a)}(u)^{-1}\bigr)},
&(2\rm a)\cr
&F_k = \prod_{j=-k+1}^{k-1}\prod_{n=0}^{k-1-\vert j \vert}
\Bigl(1+Y^{(b)}_{t_bm/t_a+j}\bigl(
u+i(k-1-\vert j \vert - 2n)/t_b\bigr)\Bigr),&(2\rm b)\cr}
$$
for $1 \le a \le r,\, m \ge 1$, where by convention
$Y^{(a)}_0(u){}^{-1} = 0$ and $Y^{(a)}_m(u) = 0\,\hbox{ if } m \not\in {\bf
Z}$.
We call (2) the (unrestricted) $\uqaxr$ functional relation.
If $Y^{(a)}_{\ell_a}(u){}^{-1} = 0$ is further imposed,
the denominator of (2a) becomes
$\prod_{j=1}^{\ell_a-1}(1+Y^{(a)}_j(u){}^{-1})^{\overline{I}_{j m}}$ with
$\overline{I}$ being the matrix $I$ of (1c) for $X_r = A_{\ell_a-1}$.
In this case, (2) closes among $\{Y^{(a)}_m(u) \vert (a,m) \in G\}$ and will be
referred as the restricted $\uqaxr$ functional relation.
The next section will concern this situation (see (6) and (15)).
Several remarks are in order:
\par\noindent
(i) The restricted FR (2) is closely related to the $\uqaxr$ Bethe equation
[18] and
its thermodynamic treatment using the string hypothesis in [8].
Under the identification
$Y^{(a)}_m(u) =$ exp$(-\beta\epsilon^{(a)}_m(u))$,
it actually corresponds to the thermal equilibrium
condition (2.22) of [8] in the high temperature limit.
\par\noindent
(ii) In the simply laced cases $X_r = A_r, D_r$ and $E_{6,7,8}$,
the restricted FR (2) takes a simple form
$$
Y^{(a)}_m(u+i)Y^{(a)}_m(u-i) =
{\prod_{b=1}^r \bigl(1+Y_m^{(b)}(u)\bigr)^{I_{a b}}
\over
\prod_{j=1}^{\ell-1}(1+Y^{(a)}_j(u){}^{-1})^{\overline{I}_{j m}}},
\eqno(3)
$$
where $\overline{I}$ denotes the $I$ for $A_{\ell-1}$.
\par\noindent
(iii) Besides the trivial $\ell = 1$ case,
(3) essentially coincides with the FR in [13,26] for $\ell = 2$
and the FR in [3,6] for $X_r = A_1$ with $\ell$ general.
The former appeared in the TBA approach to
factorized scattering theories
whilst the latter is known to hold [3,6] for certain
combinations of the actual
transfer matrices in the RSOS models [21].
In these cases, the FRs have been successfully used
to determine the central charges
by combining them with the data
from actual physical systems such as ``bulk behavior" [3],
``mass term" [9,13], etc.
\par\noindent
(iv) The FR (2) for $Y^{(a)}_m(u)$'s can be viewed as a
generalization of the recursion relation
among the quantity $Q^{(a)}_m$ in [19].
To see this, recall the definitions
$$\eqalignno{
&Q^{(a)}_m = \sum_{\bf n} Z(a,m,{\bf n})\,
\chi_{\omega(a,m,{\bf n})}, \quad\,
{\bf n} = (n_1, \ldots ,n_r) \in {\bf Z}_{\ge 0}^{\otimes r},
&(4\rm a)\cr
&\omega(a,m,{\bf n}) = m\Lambda_a - \sum_{b=1}^r n_b \alpha_b,\quad
Z(a,m,{\bf n}) = \sum_\nu \prod_{b=1}^r \prod_{k=1}^\infty
{{\cal P}^{(b)}_k(\nu) + \nu^{(b)}_k \choose \nu^{(b)}_k},&(4\rm b)\cr
&{\cal P}^{(b)}_k(\nu) = \hbox{min}(m,k)\delta_{a b} -
2\sum_{j \ge 1} \hbox{min}(k,j)\nu^{(b)}_j
- \sum_{\scriptstyle c = 1 \atop \scriptstyle c \neq b}^r \sum_{j \ge 1}
\hbox{max}(k C_{c b}, \, j C_{b c}) \nu^{(c)}_j, &(4\rm c)\cr
}$$
for $1 \le a \le r, \,\, m \ge 0$.
Here the symbol ${\quad \choose \quad}$ in (4b) is the
binomial coefficient and the sum extends over all possible
decompositions
$\{ \nu^{(b)}_k \mid\,
n_b = \sum_{k=1}^\infty k\nu^{(b)}_k, \,
\nu^{(b)}_k \in {\bf Z}_{\ge 0}, 1 \le b \le r, \, k \ge 1\}$
such that
${\cal P}^{(b)}_k(\nu) \ge 0$ for $1 \le b \le r,\, k \ge 1$.
The quantity $\chi_\omega$ in (4a) is the character of the
irreducible $\xr$ module $V_\omega$ with highest weight $\omega$, i.e.,
$
\chi_\omega = \chi_\omega(z) =
\hbox{Tr}_{V_\omega}
\hbox{ exp}\bigl(-{2\pi i\over \ell + g}(z + \rho)\bigr),
$
where we have exhibited the dependence on its argument
$z \in {\frak H}^\ast$.
We shall simply write $Q^{(a)}_m$ to mean the corresponding
$Q^{(a)}_m(z)$.
Then the following recursion relation is known
among the $Q^{(a)}_m$'s [19],
$$\eqalignno{
Q^{(a)}_m{}^2 &= Q^{(a)}_{m-1}Q^{(a)}_{m+1} +
Q^{(a)}_m{}^2 \prod_{b=1}^r\prod_{k=0}^{\infty}
Q^{(b)}_k{}^{-2J^{k\, m}_{b\, a}}\,\,\hbox{ for }
1 \le a \le r, \, m \ge 0, &(5\rm a)\cr
2J^{n\, k}_{a\, b} &= B_{a b}
\bigl({t_{a b} \over t_a}\delta_{t_b n, \, t_a k} +
{t_{a b} \over t_b}\sum_{j=1}^{t_b-t_a} j
(\delta_{t_b(n+1)-t_aj, t_a k} + \delta_{t_b(n-1)+t_aj, t_a k})
\bigr),&(5\rm b)\cr
&\qquad\hbox{ for } 1 \le a, b \le r, \quad n, k \ge 0,\cr
}$$
with the initial condition
$Q^{(a)}_{-1} = 0,\,Q^{(a)}_0 = 1$.
In (5b), the sum $\sum_{j=1}^{t_b-t_a}$  is zero unless
$t_b > t_a$ and
$J^{n\, k}_{a\, b}$ is a natural extension of the
${\hat J}^{n\, k}_{a\, b}(0)$ defined in [8]
for the range $0 < n < \ell_a, \, 0 < k < \ell_b$.
Now consider the limit $u \rightarrow \infty$ in (2).
{}From (5) the resulting algebraic equation admits a constant solution
$$
Y^{(a)}_m(\infty) =
{Q^{(a)}_m{}^2 \prod_{b=1}^r\prod_{k=0}^{\infty}
Q^{(b)}_k{}^{-2J^{k\, m}_{b\, a}}
\over
Q^{(a)}_{m+1}Q^{(a)}_{m-1}},\eqno(6)
$$
which implies that (2) is a ``$u-$version" of (5).
\beginsection 3. Dilogarithm conjecture

Here we formulate our dilogarithm conjecture firstly
then discuss its physical backgorund in the light of
the FR (2).
Let $\log x$ signify the logarithm in the branch
$-\pi < {\rm Im}(\log x) \le \pi$ for $x \neq 0$.
Consider a contour on a complex $x-$plane
${\cal C}={\cal C}[f \vert M,N], \,(f \in {\bf C},\, M,N \in {\bf Z})$
from $0$ to $f$ which intersects the branch cut of
$\log x$ for $M$ times and that of $\log(1-x)$ for $N$ times in total.
Here the intersection number is counted
as +1 when ${\cal C}$ goes across the cut of $\log x$
(resp. $\log(1-x)$) from the upper (resp. lower) half plane
to the lower (resp. upper) and as $-1$ if opposite.
Let $\Log_{\cal C}(f)$ denote the analytic continuation of $\log f$
along the contour ${\cal C}$, namely,
$$\eqalignno{
\Log_{\cal C}(f) &= \log f + 2\pi i M, \quad
\Log_{\cal C}(1-f) = \log (1-f) + 2\pi i N.&(7)\cr
}$$
Define the multivalued and single-valued Rogers dilogarithms
$L_{\cal C}(f)$ and $L(f)$ by
$$\eqalignno{
L_{\cal C}(f) &= -{1 \over 2} \int_{\cal C}
\Bigl( {\Log_{\cal C} (1-x) \over x}
     + {\Log_{\cal C}\, x \over 1 - x }\, \Bigr) dx, &(8{\rm a})\cr
L(f) &= -{1 \over 2} \int_{{\cal C}_0}
\Bigl( {\log (1-x) \over x} + { \log \, x \over 1 - x }\, \Bigr) dx,
&(8{\rm b})\cr}
$$
where ${\cal C}_0$ is a contour which does not go across the branch cuts
of $\log x$ and $\log (1-x)$.
It follows from the definitions that
$$\eqalignno{
L(f) &= \cases{
{\pi^2 \over 3}
- {\pi i \over 2} \log f - L(f^{-1}) & if $f \in {\bf R}_{>1}$,\cr
-{\pi^2 \over 6}
+ {\pi i \over 2} \log (1 - f) + L({1 \over 1 -f}) & if $f \in {\bf
R}_{<0}$,\cr}
&(9\rm a)\cr
L_{\cal C}(f) &= L(f) + \pi i M  \log (1-f) - \pi i N \log f
+ 2\pi^2 M N. &(9\rm b)\cr}
$$
\par
We shall call an element $z \in {\frak H}^\ast$ {\sl regular} if it
satisfies $Q^{(a)}_m(z) \neq
0$ for all $1 \le a \le r,\, 1 \le m \le \ell_a$ and {\sl singular} otherwise.
Suppose $z$ is regular and put
$$
f^{(a)}_m(z) =
1 - {Q^{(a)}_{m-1}(z)Q^{(a)}_{m+1}(z) \over Q^{(a)}_m(z){}^2}\quad
(a,m) \in G.\eqno(10)
$$
Then $f^{(a)}_m(z) \neq 0, 1, \infty$ for $\forall (a,m) \in G$.
Given a set of integers
${\cal S} = \{M^{(a)}_m, N^{(a)}_m \in {\bf Z} \vert (a,m) \in G \}$,
let ${\cal C}_{a,m} = {\cal C}[f^{(a)}_m(z) \vert M^{(a)}_m, N^{(a)}_m]$
be the contour as specified above.
Motivated by the study of the FR (2) which will be discussed later, we define
$$\eqalignno{
&{\pi^2 \over 6}c(z,{\cal S}) =
\sum_{(a,m) \in G}\Bigl(
L_{{\cal C}_{a,m}}\bigl(f^{(a)}_m(z)\bigr) -
{\pi i\over 2}D_{{\cal C}_{a,m}}(z)
\Log_{{\cal C}_{a,m}}\bigl(1-f^{(a)}_m(z)\bigr)\Bigr),&(11\rm a)\cr
&\pi i D_{{\cal C}_{a,m}}(z) =
\Log_{{\cal C}_{a,m}}\bigl(f^{(a)}_m(z)\bigr) -
\sum_{(b,k) \in G}A^{m\, k}_{a\, b}B_{a b}
\Log_{{\cal C}_{b,k}}\bigl(1-f^{(b)}_k(z)\bigr).&(11\rm b)\cr
}$$
By applying (7), (9b) and (11b) to (11a), one can split $c(z,{\cal S})$
into ${\cal S}-$dependent and independent parts as
$$\eqalignno{
&c(z,{\cal S}) = c_0(z) - 24T(z,{\cal S}),&(12\rm a)\cr
&{\pi^2 \over 6}c_0(z) =
\sum_{(a,m) \in G}\Bigl(
L\bigl(f^{(a)}_m(z)\bigr) -
{\pi i\over 2}d^{(a)}_m(z)
\log\bigl(1-f^{(a)}_m(z)\bigr)\Bigr),&(12\rm b)\cr
&\pi i d^{(a)}_m(z) =
\log f^{(a)}_m(z) -
\sum_{(b,k) \in G}A^{m\, k}_{a\, b}B_{a b}
\log\bigl(1-f^{(b)}_k(z)\bigr),&(12\rm c)\cr
&T(z,{\cal S}) = {1 \over 2}
\sum_{\scriptstyle (a,m) \in G \atop \scriptstyle (b,k) \in G}
A^{m\, k}_{a\, b}B_{a b} N^{(a)}_m N^{(b)}_k -
\sum_{(a,m) \in G}\bigl({1 \over 2}d^{(a)}_m(z) + M^{(a)}_m\bigr) N^{(a)}_m.
&(12\rm d)\cr}$$
Next we introduce an element $\lambda(z) \in {\frak H}^\ast$ by
$$
\lambda(z) = {1 \over 2 \pi i}\sum_{a=1}^r\sum_{b=1}^r C_{a b}
\Bigl(\sum_{j=1}^{\ell_b-1} j \log(1-f^{(b)}_j(z)) + \ell_b\bigl(
\log Q^{(b)}_{\ell_b-1}(z) - \log Q^{(b)}_{\ell_b}(z)\bigr)\Bigr)
\Lambda_a.
\eqno(13)
$$
{}From now on, we will mainly concern with the specialization
$z = \Lambda \in P_\ell$, which is relevant to our
restricted FR (2) (see (6) and (15) below).
\proclaim Conjecture.
Let $z \in {\frak H}^\ast$ be regular and put
$$\eqalignno{
&c_0(z) = {\ell{\rm dim }X_r \over \ell + g} - r -
24(\Delta^z_{\lambda(z)} + {\cal N}(z)),&(14\rm a)\cr
&\Delta^z_y = {(z \vert z + 2\rho)\over 2(\ell + g)} -
{\vert y \vert^2 \over 2\ell}
\quad {\rm for }\,\, y, z \in {\frak H}^\ast.&(14\rm b)\cr
}$$
Suppose $\Lambda \in P_\ell$.
Then ${\cal N}(\Lambda) \in {\bf Z}$ if $\Lambda$ is regular.
In case $\Lambda$ is singular, ${\cal N}(z)$ converges to finitely many
integers depending on the ways $z \in {\frak H}^\ast_{\bf R}$
approaches $\Lambda$.

This conjecture has been supported by numerical experiments for
$X_r = A_r, B_r$, $C_r$ and $D_r$ with small values of
the level $\ell$ and rank $r$.
It is not difficult to prove ${\cal N}(z) \in {\bf R}$ for
any regular $z \in {\frak H}^\ast_{\bf R}$.
Because of the discontinuity of the $\log$ function,
$c_0(z),\, \lambda(z)$ and therefore
${\cal N}(z)$ assume finitely many values when
$z$ approaches a singular $\Lambda$.
Hereafter we shall arbitrarily fix one way of letting
$z \rightarrow \Lambda\, (z: \hbox{regular},\, z \in {\frak H}^\ast_{\bf R})$
and all the formulas involving $\Lambda \in P_\ell$ should
be understood as defined by
this limit whenever $\Lambda$ is singular.
{}From numerical tests,
$Q^{(a)}_m(\Lambda) = Q^{(a)}_{\ell_a}(\Lambda)
Q^{(a)}_{\ell_a-m}(\Lambda)^\ast$ seems valid for any
$-1 \le m \le \ell_a+1$ and $\Lambda \in P_\ell$,
where $\ast$ denotes complex conjugation.
Note in particular that it implies
$$Q^{(a)}_{\ell_a+1}(\Lambda) = 0. \eqno(15)$$
Assuming this and (A.10,11) in [8], one can show that
$\lambda(\Lambda)$ is finite and satisfies
$$\eqalignno{
\Lambda &\equiv \lambda(\Lambda) \quad \hbox{ mod } Q,&(16\rm a)\cr
T(\Lambda,{\cal S})
&= -{1 \over 2\ell}\vert \lambda(\Lambda) + \beta({\cal S}) \vert^2 +
{1 \over 2\ell}\vert \lambda(\Lambda)\vert^2
+ \hbox{integer},&(16\rm b)\cr
\beta({\cal S}) &= \sum_{a=1}^r\sum_{m=1}^{\ell_a-1} m N^{(a)}_m \alpha_a
\in Q,&(16\rm c)\cr}
$$
where the integer part is dependent on $\Lambda$ and ${\cal S}$.
Combining (16b) with (12a) and admitting the conjecture, we have
$$
c(\Lambda,{\cal S}) =
{\ell{\rm dim }X_r \over \ell + g} - r -
24(\Delta^\Lambda_{\lambda(\Lambda) + \beta({\cal S})}
+ \hbox{integer}) \quad \hbox{for any } \, \Lambda \in P_\ell.
\eqno(17)
$$
The RHS of (17) with the congruence
properties (16a,c) is well known as the
central charge $-$ 24(scaling dimension mod ${\bf Z}$) of the
level $\ell \, \axr$ parafermion CFT [22].
As the integer set ${\cal S}$ is chosen variously, $\beta({\cal S})$ (16c)
ranges over the root lattice $Q$.
Thus all the spectra in the parafermion CFTs seem to come out
from the Rogers dilogarithm function through the quantity
$c(\Lambda,{\cal S})$ (11a).
This is our main observation in this Letter.
We note that in the case $\Lambda = 0 \in P_\ell$, $\lambda(0) = 0$
holds and the earlier conjecture in [20,8] corresponds to
${\cal N}(0) = 0$.
\par
There are actually two sources for considering the
quantity $c(\Lambda,{\cal S})$ (11a) in connection to the
parafermion CFTs, which we shall now explain.
The physical meaning of the integer set ${\cal S}$ is yet to be clarified
in the light of these connections.
The first source is to calculate finite-size corrections
to the ground state energies for $\axr$ fusion RSOS models.
In [3] Kl\"umper and Pearce evaluated them for $\axr = A^{(1)}_1$
via the Rogers dilogarithm and showed that
the resulting exponents in regime I/II become the parafermion values
in agreement with the earlier result in [21].
They started from the FR
essentially identical to the
restricted $U_q(A^{(1)}_1)$ FR (2), where
$Y^{(1)}_m(u)$ indeed is the finite-size correction
part of the row-to-row transfer matrices.
Such an interpretation is not known so far for the
$\uqaxr$ FR (2) in general.
However we have found that the $c(\Lambda,{\cal S})$ naturally arises
as the ``$c$ for excitations" in their sense
through a heuristic calculation extending the $A^{(1)}_1$ case.
Although the treatment in [3] is not necessarily identical to our
formulation here, the regime I/II result therein effectively
confirms our conjecture for the simplest case $\axr = A^{(1)}_1$.
We shall emphasize that for higher rank algebras,
the present $c(\Lambda,{\cal S})$ only corresponds to
``regime I/II-like region" with a special
``fusion type $\forall s_a > 1$" in the sense of
section 4.2 in [8].
Thus we expect further generalizations should be possible related to
various coset CFTs.
See for example the regime III/IV result in [3] for $A^{(1)}_1$.
\par
The second source is to study integrable perturbations [14] of parafermion
CFTs.
In [9-13], ultraviolet Casimir energies in various
perturbed CFTs have been obtained by using
integral equations of the form
$$
Rm^{(a)}_j \hbox{ch } u = \epsilon^{(a)}_j(u)
+ \sum_{(b,k) \in G} \int_{- \infty}^\infty
dv \, \Psi^{j k}_{a b}(u - v)\, \log
\bigl(1 + \hbox{exp}(-\epsilon^{(b)}_k(v)) \bigr),\eqno(18)
$$
for $(a,m) \in G$, which originates in the TBA.
Here $m^{(a)}_j$ denotes mass,
$\epsilon^{(a)}_j(u)$ is the energy of the physical
excitation with rapidity $u$,
$\Psi^{j k}_{a b}(u)$ is some kernel and
$R$ is the system size corresponding to the inverse temperature in the TBA.
In addition, there are some observations that
(18) also yields low-lying excitation energies
by introducing ``imaginary chemical potentials" [23,27,28].
Our $c(\Lambda,{\cal S})$ stems from a modification of (18) so as to
match the logarithm of (2) under the identification
$Y^{(a)}_j(u) = e^{-\epsilon^{(a)}_j(u)}$
up to the LHS which is tending to zero
in the ultraviolet (or high temperature) limit.
In particular, we choose $\forall m^{(a)}_j>0$ as $j-$independent, add
the imaginary term $\pi i D_{{\cal C}_a,j}(\Lambda)$ (11b) on the RHS and take
$\Psi^{j k}_{a b}(u)$ to
be the universal kernel occurring in the $\uqaxr$ Bethe equations, i.e.,
$2\pi\Psi^{j k}_{a b}(u) = \int dx e^{iux} (\delta_{a b}\delta_{j k} -
{\hat M}_{a b}(\pi x/2){\hat A}^{(\ell)\, j k}_{\quad\, a b}(\pi x/2))$
using (2.10) of [8].
Then an analogous calculation to [11] amounts to considering
the $c(\Lambda,{\cal S})$ as the ``$c$ for excitations".
\par
These observations
have led to the quantity
$c(\Lambda,{\cal S})$ which has been conjectured to possess
the remarkable property (17).
Our $\uqaxr$ FR (2) lies as their common background.
It is to be analyzed further
to actually yield the physical
consequences in the $\uqaxr$ Bethe ansatz systems.
The details omitted here will appear elsewhere.
\par \vskip0.1cm
The authors thank P.A.Pearce for supporting
their visit to University of Melbourne, kind hospitality and
explaining the work [3].
They also thank M.T.Batchelor, R.J.Baxter, V.V.Bazhanov, J.Suzuki,
M.Wakimoto and R.B.Zhang
for valuable comments.
This work is supported by the Australian Research Council.
\beginsection References

\noindent
\item{[1]}{R.J.Baxter, Physica{\bf 18D} (1986) 321}
\item{[2]}{A.Kl\"umper, M.T.Batchelor and P.A.Pearce, J.Phys.A{\bf 24} (1991)
3111}
\item{[3]}{A.Kl\"umper and P.A.Pearce, ``Conformal weights of RSOS lattice
models
and their fusion hierarchies", preprint No.23-1991 (1991)}
\item{[4]}{I.Affleck, Phys.Rev.Lett.{\bf 56} (1986) 746}
\item{[5]}{H.M.Babujan, Nucl.Phys.{\bf B215}[FS7] (1983) 317;}
\item{}{H.J.de Vega and M.Karowski, Nucl.Phys.{\bf B285}[FS19] (1987) 619;}
\item{}{A.N.Kirillov and N.Yu.Reshetikhin, J.Phys.A{\bf 20} (1987) 1587}
\item{[6]}{V.V.Bazhanov and Yu.N.Reshetikhin, Int.J.Mod.Phys.A{\bf 4}
(1989) 115}
\item{[7]}{V.V.Bazhanov and Yu.N.Reshetikhin, J.Phys.A{\bf 23} (1990) 1477}
\item{[8]}{A.Kuniba, ``Thermodynamics of the $\uqaxr$ Bethe Ansatz System
With $q$ A Root of Unity", ANU preprint (1991)}
\item{[9]}{Al.B.Zamolodchikov, Nucl.Phys.{\bf B342} (1990) 695;
{\bf B358} (1991) 497}
\item{[10]}
{V.V.Bazhanov and N.Yu.Reshetikhin, Prog.Theor.Phys.Suppl.{\bf 102} (1990) 301}
\item{[11]}{M.J.Martins, Phys.Rev.Lett.{\bf 65} (1990) 2091}
\item{[12]}{T.R.Klassen and E.Melzer, Nucl.Phys.{\bf B338} (1990) 485;
{\bf B350} (1991) 635}
\item{[13]}{V.A.Fateev and Al.B.Zamolodchikov, Phys.Lett.{\bf B271} (1991) 91}
\item{[14]}{A.B.Zamolodchikov, Int.J.Mod.Phys.{\bf A4} (1989) 4235}
\item{[15]}{H.W.J.Bl\"ote, J.L.Cardy
and M.P.Nightingale,  Phys.Rev.Lett.{\bf 56} (1986) 742}
\item{[16]}{L.Lewin, {\sl Polylogarithms and associated functions},
(North-Holland, 1981)}
\item{[17]}{C.N.Yang and C.P.Yang, J.Math.Phys.{\bf 10} (1969) 1115}
\item{[18]}
{N.Yu.Reshetikhin and P. B. Wiegmann, Phys.Lett{\bf B189} (1987) 125}
\item{[19]}{A.N.Kirillov and N.Yu.Reshetikhin,
Zap.Nauch.Semin.LOMI{\bf 160} (1987) 211}
\item{[20]}{A.N.Kirillov, Zap.Nauch.Semin.LOMI{\bf 164} (1987) 121 and private
communications}
\item{[21]}
{G.E.Andrews, R.J.Baxter and P.J.Forrester,
J.Stat.Phys.{\bf 35} (1984) 193;}
\item{}{E.Date, M.Jimbo, A.Kuniba, T.Miwa and M.Okado,
Nucl.Phys.{\bf B290}}
\item{}{[FS20] (1987) 231; Adv.Stud.in Pure Math.{\bf 16}
(1988) 17}
\item{[22]}{V.A.Fateev and A.B.Zamolodchikov,
Sov.Phys.JETP {\bf 62} (1985) 215;}
\item{}{D.Gepner, Nucl.Phys.{\bf B290}[FS20] (1987) 10}
\item{[23]}{T.R.Klassen and E.Melzer, Nucl.Phys.{\bf B370} (1992) 511}
\item{[24]}{V.G.Kac, {\sl Infinite dimensional Lie algebras},
(Cambridge University Press, 1990)}
\item{[25]}{M.Jimbo, Lett.Math.Phys.{\bf 10} (1985) 63;}
\item{}{V.G.Drinfel'd, ICM proceedings, Berkeley (1987) 798}
\item{[26]}{Al.B.Zamolodchikov, Phys.Lett.{\bf B253} (1991) 391}
\item{[27]}{M.J.Martins, Phys.Rev.Lett.{\bf 67} (1991) 419}
\item{[28]}{P.Fendley, Nucl.Phys.{\bf B372} (1992) 533}
\bye